# Modeling the dynamics of inhomogeneous natural rotifer populations under toxicant exposure

Georgy P. Karev [a], Artem S. Novozhilov [a,], and Faina S. Berezovskaya [b]

[a] *National Institutes of Health, National Center for Biotechnology Information, National Library of Medicine, Bethesda, MD 20894, USA*

[b] *Department of Mathematics, Howard University, Washington, DC 20059, USA*

**Abstract**

Most population models assume that individuals within a given population are identical, that is, the fundamental role of variation is ignored. Here we develop a general approach to modeling heterogeneous populations with discrete evolutionary time step. The theory is applied to population dynamics of natural rotifer populations. We show that under particular conditions the behavior of the inhomogeneous model possesses complex transition regimes, which depends both on the mean and the variance of the initial parameter distribution and the final state of the population depends on the least possible value from the domain of the parameter. The question of persistence of the population is discussed.

## 1. Introduction

Due to ecological importance of zooplankton, a substantial amount of data exists on the abundance of natural populations in lakes, estuaries and coastal marine environments. A variety of mathematical models have been applied for modeling the dynamics of zooplankton populations (e.g., McCauley et al., 1996; Snell and Serra, 1998). Recently a particular class of mathematical models, extracted as deterministic dynamics components from noisy ecological time series was studied systematically (Berezovskaya et al., 2005). This class of models was primarily developed to analyze the dynamics of natural rotifer populations and evaluate the ecological consequences of toxicant exposure.

Modeling population dynamics often involves balancing the competing requirements of realism and simplicity. On the side of simplicity, one has classical population models with discrete time, which have been extensively studied for decades (e.g., pioneer works of Shapiro (1973) and May (1975), among many others). It is a well-known fact that these models can possess various dynamical behaviors varying from stable steady states and cycles to chaotic oscillations. These models keep track of the total population size and treat all individuals as identical. That is, the fundamental role of variation is ignored, and parameter values represent some averaged value whilst the information on variance and other characteristics are not taken into account. This, of course, simplifies the computation at the cost of realism. In recent years, several researchers have focused on generalizing continuous time population models in such a way to allow for different individuals or subpopulations to have different growth or mortality rates (Karev, 2000, 2005; more abstract approach was developed earlier in works of Semenov, Okhonin, Gorban, see survey of Gorban (2005) for details and references). It was shown that recognition of heterogeneity may lead to unexpected and even counter-intuitive effects. Here we present a general framework to analyze the dynamics of inhomogeneous populations with discrete time steps, and apply it to the study of dynamical behavior of heterogeneous rotifer populations.

The paper is organized as follows. In Sec. 2 we summarize main results from Berezovskaya et al. (2005), which include parametric portrait of the model for population dynamics of rotifer populations, and discuss questions on population persistence and extinction. In Sec. 3 we formulate general framework for analyzing heterogeneous models with discrete time. In Sec. 4 we apply the theory of inhomogeneous maps to the model of population dynamics of rotifer populations. In particular, we show that the knowledge of mean values of parameters is not sufficient for predicting the evolution and eventual fate of the population. Moreover, the behavior of populations with the same initial mean values of parameters and different variances can differ dramatically: one of the populations can go extinct while other can reach a stable stationary regime. Section 5 is devoted to discussion and conclusions.

## 2. The Consensus model

Methods developed to extract deterministic dynamics components from short noisy time series (Aksakaya et al., 1999) were applied to data from natural populations of nine rotifer species (Snell and Serra, 1998). Time series of a population density $N_t$ (a number of organisms per liter at time $t$ with time unit equals to 2 days) have been received. Using these series several phenomenological models have been checked for fitting. The best-fit model for 5 of 9 data sets (named the Consensus model) has the following scaled form:

$$N_{t+1} = N_t \exp\{-a + 1/N_t - \gamma/N_t^2\}, \tag{1}$$

where $a \geq 0$ is the parameter characterizing density-independent effects on the reproduction rate, which can be interpreted as an environmental press to rotifers (poor water quality, extreme temperature or toxicant exposure), and $\gamma$ is the species-specific parameter. Berezovskaya et al. (2005) showed that depending on parameter values the asymptotic behavior of (1) can vary from equilibrium points to chaotic oscillations with usual period-doubling route to chaos. Moreover, model (1) possesses the property of bistability (strong Allee effect [e.g., Wang, and Kot, 2001]). This means that there exists a threshold level of population size such that if the total size of the population is less than this quantity the extinction of the population is certain. The diversity of behaviors of model (1) is associated with the complex form of the map. The main results of the analysis of (1) are summarized in its parametric portrait (Fig. 1a).

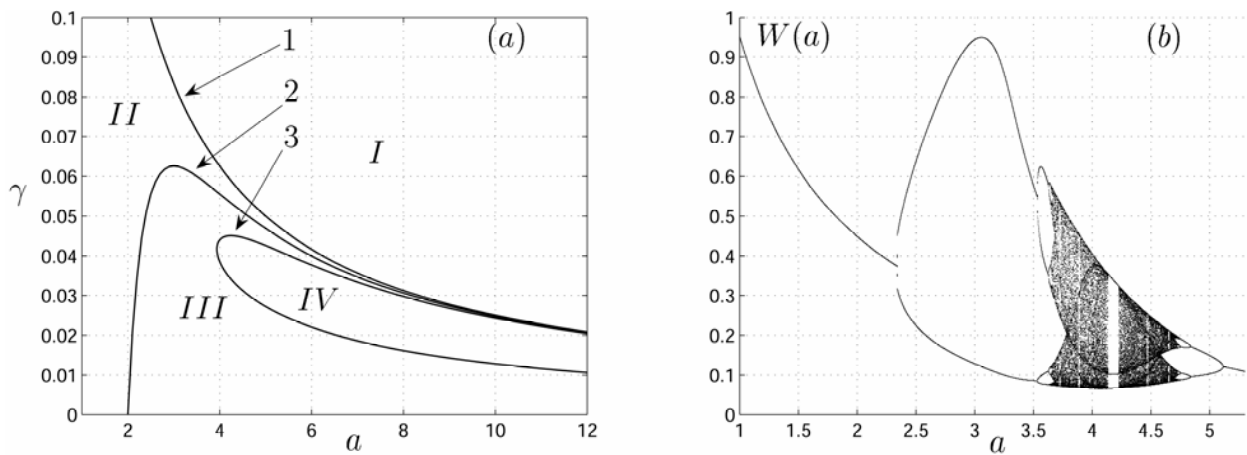

Fig. 1. (a) Parametric portrait of model (1). The boundaries of the domains: $\gamma = 0$; 1 - $\{\gamma \,|\, \gamma = 1/(4a)\}$; 2 - $\{\gamma \,|\, \gamma = (a-2)/(4(a-1)^2)\}$; 3 – the boundary of the 0-attracting domain. The domains are I – Total extinction; II – Bistability; III – Oscillations and chaos and zero stable; IV – Total extinction through aperiodic oscillations, zero stable (the 0-attracting domain). (b) Bifurcation diagram of (1) for $\gamma = 0.046$. $W(a)$ is the set of observed states of the population for the given parameter values

In Fig. 1a there are two domains (II and III) where persistence of the population is possible, though it should be noted that zero is a stable stationary point in these domains, i.e., if the total size of the population is lower that some threshold level, the extinction is certain. In domains I and IV any initial values of the population size lead to zero stable state, the population cannot survive. Different possible ways of population extinction if the parameters are varied were analyzed, see (Berezovskaya et al., 2005) for details. What is important, model (1) has a density independent parameter $a$, which can be also interpreted as an average ability of an individual to reproduce under given toxicant exposure (compare with the above interpretation). If the population is highly heterogeneous, i.e., there are some individuals that can stand and reproduce under the conditions, and there are some for which pollution is mortal, one has to adjust the model in such a way to allow for different individuals have different reproduction rates.

### 3. Inhomogeneous maps theory

Let us assume that a population consists of individuals, each of those is characterized by its own parameter value $\mathbf{a} = (a_1,...,a_k)$. These parameter values can take any particular value from set A. Let $n_t(a)$ be the density of the population at the moment $t$. Then the number of individuals having parameter values in set $\tilde{A} \subseteq A$ is given by $\tilde{N}_t = \int_{\tilde{A}} n_t(\mathbf{a}) d\mathbf{a}$, and the total population size is $N_t = \int_A n_t(\mathbf{a}) d\mathbf{a}$.

If a model with discrete time steps is applicable to the population, then in the next time instant $t+1$, we should have $n_{t+1}(\mathbf{a}) = W n_t(\mathbf{a})$, where $W \geq 0$ is the reproduction rate. In what follows we assume that the reproduction rate depends on the specific parameter value $\mathbf{a}$, and the total size of the population, i.e., $W = W(N_t, \mathbf{a})$. It means that the population dynamics is governed by the model of the form

$$n_{t+1}(\mathbf{a}) = W(N_t, \mathbf{a}) n_t(\mathbf{a}), \qquad N_t = \int_A n_t(\mathbf{a}) d\mathbf{a}. \tag{2}$$

The initial distribution $n_0(\mathbf{a})$ is supposed to be given.

Let us denote $p_t(\mathbf{a}) = n_t(\mathbf{a}) / N_t$ the current probability density function (pdf) of the vector-parameter $\mathbf{a}$ at the moment $t$. We have a probability space $(A, p_t(\mathbf{a}))$, and model (2) defines a transformation of the initial pdf $p_0(\mathbf{a})$ with time. Below we show that problem (2) can be reduced to a non-autonomous map on $I \subseteq \mathbf{R}^1$ under supposition that the reproduction rate has the form $W(N,\mathbf{a}) = f(\mathbf{a})g(N)$, so that the model takes the form

$$n_{t+1}(\mathbf{a}) = n_t(\mathbf{a})f(\mathbf{a})g(N_t), \qquad N_t = \int_A n_t(\mathbf{a})d\mathbf{a}, \qquad (3)$$

for the given initial density $n_0(\mathbf{a})$.

Rewriting the first equation in (3) as $n_{t+1}(\mathbf{a})/n_t(\mathbf{a}) = f(\mathbf{a})g(N_t)$, we obtain

$$n_t(\mathbf{a}) = n_0(\mathbf{a})f^t(\mathbf{a})G_{t-1}, \qquad (4)$$

where $G_t = g(N_0) \cdot ... \cdot g(N_t)$. Then, using (4),

$$N_t = \int_A n_t(\mathbf{a})d\mathbf{a} = \int_A n_0(\mathbf{a})f^t(\mathbf{a})G_{t-1}d\mathbf{a} = N_0 \mathrm{E}_0[f^t]G_{t-1}, \qquad (5)$$

where $\mathrm{E}_0[f^k] = \int_A f^k(\mathbf{a})p_0(\mathbf{a})d\mathbf{a}$ is the $k$-th moment of the initial distribution of $f(\mathbf{a})$. From (4) and (5) one has

$$p_t(\mathbf{a}) = \frac{n_t(\mathbf{a})}{N_t} = p_0(\mathbf{a})\frac{f^t(\mathbf{a})}{\mathrm{E}_0[f^t]}. \qquad (6)$$

Integrating over $\mathbf{a}$ the equation $n_{t+1}(\mathbf{a}) = f(\mathbf{a})p_t(\mathbf{a})N_t g(N_t)$ implies

$$N_{t+1} = \mathrm{E}_t[f]N_t g(N_t),$$

where $\mathrm{E}_t[f] = \int_A f(\mathbf{a})p_t(\mathbf{a})d\mathbf{a}$ is the mean value of $f$ at $t$ moment.

Next, according to (6),

$$\mathrm{E}_t[f] = \frac{1}{E_0[f^t]} \int_A f^{t+1}(\mathbf{a})p_0(\mathbf{a})d\mathbf{a} = \frac{E_0[f^{t+1}]}{E_0[f^t]}, \qquad (7)$$

and finally we can state the following basic theorem.

**Theorem 1.** *Let $p_0(\mathbf{a})$ be the initial distribution of the vector-parameter $\mathbf{a}$ for inhomogeneous map* (2). *Then*

i) *The population size $N_t$ satisfies the recurrence relation $N_{t+1} = \mathrm{E}_t[f]N_t g(N_t)$;*

ii) *The current mean of $f$ can be computed by the formula $\mathrm{E}_t[f] = E_0[f^{t+1}]/E_0[f^t]$;*

iii) *The current pdf is given by the formula $p_t(\mathbf{a}) = p_0(\mathbf{a})f^t(\mathbf{a})/\mathrm{E}_0[f^t]$.*

## 4. Simulation analysis of inhomogeneous rotifer population dynamics

Using model (1) as a prototype equation which governs the dynamics of a rotifer population and assuming that parameter $a$ is distributed among individuals of the population, we have the model in the form (3) with $\mathbf{a} = a$, $f(\mathbf{a}) = \exp\{-a\}$, and $g(N) = \exp\{1/N - \gamma/N^2\}$. In biological terms, the distribution of $a$ can be interpreted as different ability of reproduction for different individuals of the population under constant toxicant exposure. Due to the interplay of two independent factors – reproduction ability and density dependent population regulation – the structure of the population will be evolving in such a way that fittest individuals will be replacing less fit ones. On the other hand, in the considered case, the population regulation involves the strong Allee effect, which means that under particular circumstances it is possible for the total size of the population to fall below the threshold level, which means eventual extinction of the population in spite of the fact that some of the individuals might perfectly exist under given pollution level.

To illustrate the possible dynamical behavior of a rotifer population we must have an initial distribution of $a$, which is unknown. There can be several reasonable choices. First, note, that $a \geq 0$, which means that we must confine ourselves to nonnegative distributions. It is natural to choose a unimodal initial distribution, and in this case, an obvious candidate for the initial distribution is a Gamma-distribution since it can well approximate almost any unimodal distribution concentrated on the positive half-line. Other possible distributions include uniform, log-normal, Pareto, and many others. Evolution of all these distribution is described by Theorem 1,iii. Other approaches to account for population heterogeneity are discussed below.

Let us assume that the initial distribution of $a$ is a Gamma-distribution with parameters $b, k, s$, $p_0(a) = \dfrac{s^k}{\Gamma(k)}(a-b)^{k-1}\exp\{-(a-b)s\}$, where $\Gamma(k)$ is the gamma-function, $a \geq b$, $k, s > 0$. We have $\mathrm{E}_0[a] = b + k/s$, $\mathrm{Var}_0[a] = k/s^2$. If $k = 0$ then we deal with an exponential distribution on $[b,\infty)$. Straightforward calculations show that $E_0[f^t] = \exp\{-bt\}s^k/(s+t)^k$. From Theorem 1,iii, it follows that $p_t(a)$ is the Gamma-distribution with parameters $b, k, s+t$; $\mathrm{E}_t[a] \to b,\ \mathrm{Var}_t[a] \to 0$ when $t \to \infty$; from Theorem 1,ii one obtains

$E_t[f] = \exp\{-b\}(s+t)^k / (s+t+1)^k$, and from Theorem 1,i the dynamics of the total population size is governed by the recurrence relation

$$N_{t+1} = N_t \exp\{-b\} \left[\frac{s+t}{s+t+1}\right]^k \exp\left\{\frac{1}{N_t} - \frac{\gamma}{N_t^2}\right\}. \tag{8}$$

From (8) it is seen that formally the evolution of the heterogeneous population is accompanied by replacing individuals with higher values of $a$ by individuals with lower values of the parameter. It agrees with the movement in the parameter space of (1) (Fig. 1a) with fixed $\gamma$ and monotonically decreasing $a$ to its minimal possible value (equals $b$ in the considered case), which determines the asymptotical state of the population in the case of no extinction. What matters is the speed with which the quantity $E_t[f]$ changes with time. In Fig. 2 $E_t[f]$ is plotted with the same initial mean and different variances of $a$. From Fig. 2 it is clear that the smaller the initial variance the slower the rate of reaching the limiting value. Moreover, note that for some particular values of the parameters of the initial distribution, the function $E_t[f]$ has an inflection point at $t = (k-1)/2 - s$.

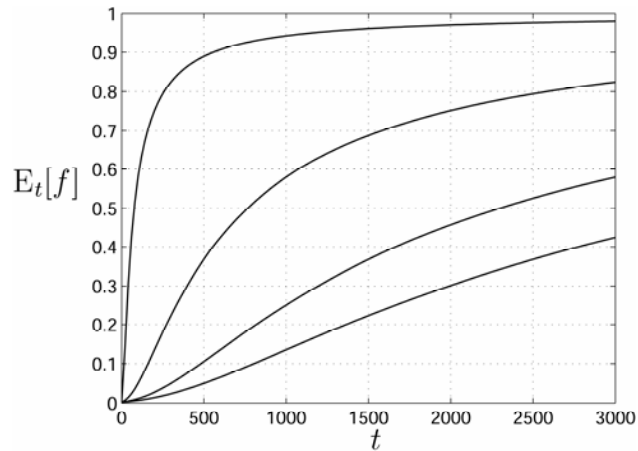

Fig. 2. $E_t[f]$ versus time in the case of Gamma-distribution of $a$: $E_0[a] = 6$, $Var_0[a] = 0.6, 0.06, 0.02, 0.012$ from top to down curves; $b = 0$

Now let us assume that an isolated rotifer population is exposed to some constant environmental press (e.g., the water is poisoned with some pollutant). Taking into account distribution of parameter $a$ we assume that some individuals are better adapted to the situation and some cannot survive. Natural selection and density dependent population regulation lead to changing of the structure of population. To illustrate possible dynamical behavior of the population we fix the initial mean of the distribution and consider a number of different initial variances. In all the simulations depicted we take $b = 0$ (i.e., $E[a] \to 0$ when $t \to \infty$). The results are shown in Fig. 3.

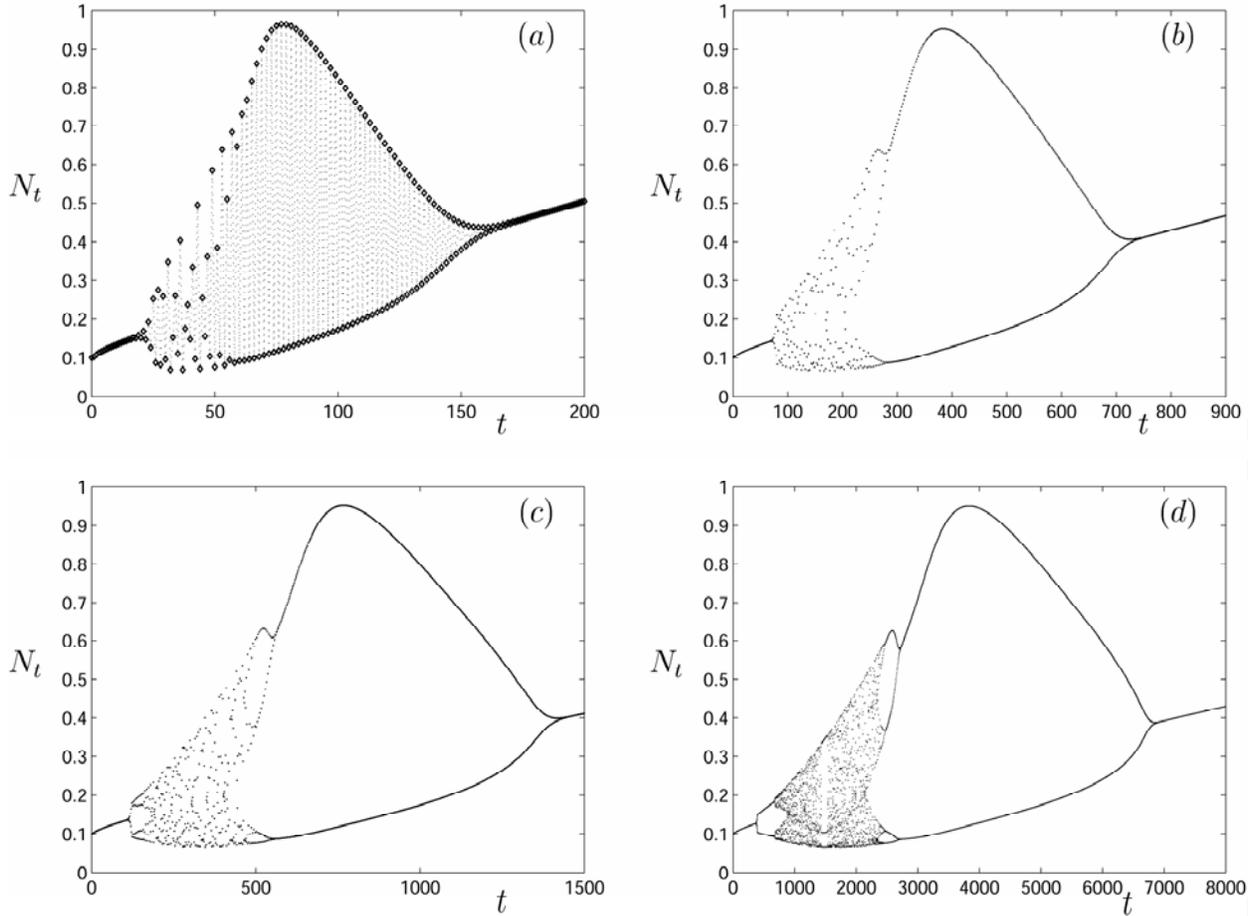

Fig. 3. The behavior of total population size $N_t$ in inhomogeneous model (8). The path is depicted only in (a). The initial mean $E_0[a] = 5.4$ for all four cases. The initial variance $Var_0[a]$ is (a) 0.054, (b) 0.0108, (c) 0.0054, (d) 0.00108. The value of $b$ in Gamma distribution is taken zero, and $\gamma = 0.046$.

Fig. 3. allows us to make several conclusions. First, we take such value of $\gamma$ that the trajectory we follow during simulation in the parametric space of (1) lies above the 0-attraction domain (Fig. 1a). The choice of $\gamma$ reflects the fact that our primary goal is to cross a maximal number of different domains in Fig. 1a during simulation. The bifurcation diagram of (1) for the given $\gamma$ is represented in Fig. 1b. Second, the lower the initial variance of the distributed parameter (i.e., population is less heterogeneous) the more time is needed to reach the asymptotical state. We indicate here that we use $b = 0$ in our simulations, which means that the asymptotical state of the population (when $t \to \infty$) is an unbounded growth. This is an unrealistic assumption, but all the calculations can be done in the same manner with $b > 0$, and if one takes $b < 1$ then the pictures do not change qualitatively. The only thing that should be taken into account is the fact that the speed of reaching the final state becomes slow as soon as the population is close to this asymptotical state (see Fig. 2).

Third, the trajectory $\{N_t\}_0^\infty$ in some sense mimics the bifurcation diagram in Fig. 1b (note that larger values of $t$ correspond to smaller values of $a$ in the bifurcation diagram, that is, to compare the two figures one should reverse one of them), but the bifurcation diagram cannot be used (with formal redefinition of parameters) to infer a trajectory of the inhomogeneous model. When drawing the bifurcation diagram one has to assume that the parameters are constant, whereas plotting trajectories of the inhomogeneous model (8) we formally consider the system with varying parameter. Dynamical systems having parameters that change in time (no matter how slowly) and pass through bifurcation values often exhibits behavior that is very different from the analogous situation where the parameters are constant (Wiggins, 1990, pp. 384-386, and references therein). A usual picture in such cases is a delay in bifurcation (Arnold, 1990). For example, the flip bifurcation of (1) with $\gamma = 0.046$ happens when $a = 2.32$ (birth of a stable 2-cycle if one goes from left to right in Fig.1a) and $a = 5.11$ (disappearance of a stable 2-cycle if one goes from left to right in Fig. 1a), which corresponds to $t = 132$ and $t = 5$ in the case of $E_0[a] = 5.4$, $Var_0[a] = 0.054$. This is clearly not the case (see Fig. 3a). The same situation is observed with other possible parameter values.

Forth, on its way to a stable state the total size of the population can experience dramatically different behavior from apparently chaotic oscillations to oscillatory-like changes to smooth changes in population size (Fig. 3). This fact is the consequence of interplay of two independent factors: heterogeneity of the population and density-dependent regulatory mechanism. We would like to emphasize that the evolution of the distribution of the parameter (that is, the behavior of frequencies of different type individuals) is regular and completely described by Theorem 1. It can be proven that in many important cases the distribution of parameters is of the same "type" as the initial one (i.e., Gamma or Beta –distributions), but with changing in time parameters.

Since population regulation can be unfavorable not only for a large number of individuals, but also for small population (which is the case in model (1)), conclusions based on the frequency analysis can be misleading. Let us suppose that $\gamma = 0.044$ and $E_0[a] = 5.2$, i.e. we start from the point in the parametric space of (1) (Fig. 1a) such that on the way to smaller values of $a$ the trajectory has to cross the 0-attracting area. As numerical simulations show, depending on the initial state of the population, there is a threshold value of $Var_0[a]$ such that for larger variances the population jumps over the extinction area, and the evolution of the population is similar to the cases considered above (Fig. 3 and 4a), but for smaller variances the population is trapped in the 0-attraction area and goes extinct (Fig. 4b). In spite of the fact that some of the individuals of the population can perfectly survive under given toxicant exposure, the law of population regulation (existence of threshold size) prevents survival of the whole population.

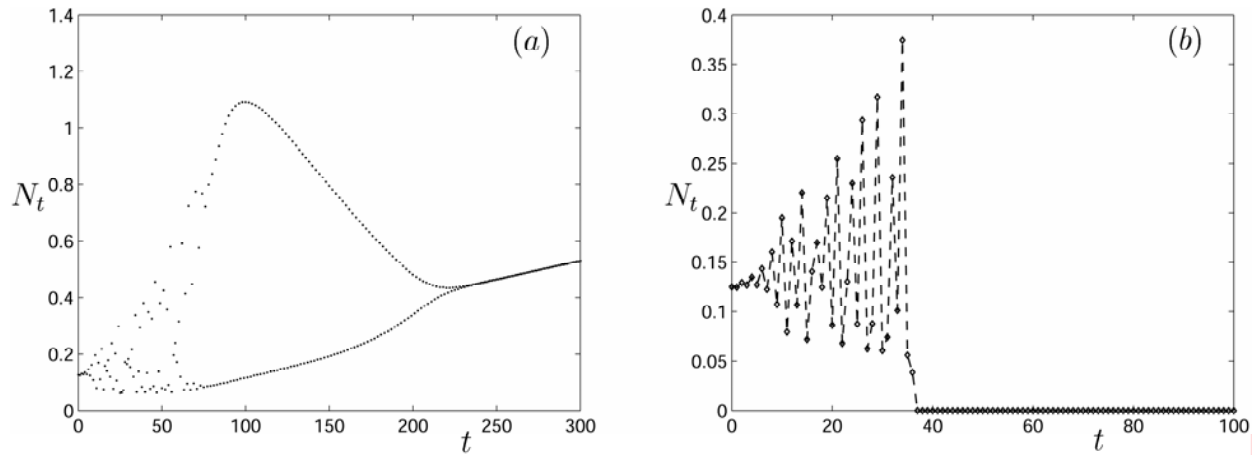

Fig. 4. The behavior of total population size $N_t$ in inhomogeneous model (8). $\gamma = 0.044$, $E_0[a] = 5.2$. (a) The initial variance is $Var_0[a] = 0.035$ and the population reaches the stable asymptotical state. (b) The initial variance is $Var_0[a] = 0.026$ and the population goes extinct being trapped in the 0-attracting domain

In our simulations we used a Gamma-distribution as an initial distribution of the parameter $a$. In most real situations the initial distribution is obviously unknown and cannot be obtained with high accuracy. Our choice was stipulated by the fact that Gamma-distribution can approximate any unimodal distribution. There are other possible choices. Noting that $f(a)$ in (1) belongs to the interval $(0,1]$, we could assume an initial distribution for $f$. A wide class of distributions with bounded support is Beta-distributions. Using Theorem 1, *iii*) one can prove the if $p_0(f)$ has a Beta-distribution in $(0, 1)$ with parameters $\alpha, \beta$, then $p_t(f)$ is again Beta-distribution with parameters $\alpha+t, \beta$, and $E_t[f]=(\alpha+t)/(\alpha+\beta+t)$. Qualitatively the behavior of the model with a beta-distributed parameter is similar to the behavior of the model with a gamma-distributed parameter, the only difference is the speed to reach the final state of the population.

## 5. Discussion

In the paper we presented a unified approach to model heterogeneous populations when they have a distinct discrete evolutionary time step. The general theory was illustrated and highlighted by simulation analysis of heterogeneous model of rotifer populations. We showed that the knowledge of an average parameter value is not sufficient for predicting the eventual fate of the population under consideration, and other characteristics such as variance should be taken into account for thorough modeling purposes.

Asymptotically, the inhomogeneous model behavior tends to that of homogeneous model with the parameter *a* equal to the least possible value from the domain of *a*. It reflects the fact that, inasmuch as in our model we do not consider mutations, the evolution of distribution corresponds to the simplest case of the Haldane principle, when more fit individuals replace less fit individuals (Haldane, 1990; see also Karev (2005) for specific form of this principle for inhomogeneous population).

The main peculiarity of the inhomogeneous model compared to the homogeneous one, is the existence of the transition behavior. The transition to the asymptotical regime could be very complex, see Fig. 3 and 4. The mean value $E_t[a]$ of parameter *a* changes with time (and moves toward the least possible value from the domain of *a*); the step $\Delta E_t = E_{t+1}[a] - E_t[a]$ tends to 0 at large *t*. At small values of the step $\Delta E_t$ the model orbit mimics a subset of the bifurcation diagram of the initial homogeneous model; this remark helps to understand why the transition regimes might be very complex. At small *t* the value of step $\Delta E_t$ depends on the model parameters, in particular on the variance of the initial parameter distribution (the smaller the variance the smaller the step). At relatively large variances the system quickly transits to an asymptotical regime; and if the variance is small enough, this transition may take a long time and can be dangerous: on the way to a possible stable asymptotical regime the population can enter the 0-attracting domain which means extinction (Fig. 1a and 4b).

Acknowledgment. The work of F. Berezovskaya was supported by NSF Grant #634156.